\documentclass[letterpaper]{article} 
\usepackage{aaai2027}  
\usepackage[hyphens]{url}  
\usepackage{graphicx} 
\usepackage{natbib}  
\usepackage{caption} 
\usepackage{amsmath}
\usepackage{amssymb}
\usepackage{algorithm}
\usepackage{algpseudocode}
\usepackage{multirow} 
\usepackage{makecell}

\usepackage{newfloat}
\usepackage{listings}
\DeclareCaptionStyle{ruled}{labelfont=normalfont,labelsep=colon,strut=off} 
\floatstyle{ruled}
\newfloat{listing}{tb}{lst}{}
\floatname{listing}{Listing}

\usepackage{booktabs}

\title{Reliability Testing of Medical Model Performance under Distributed Deployment}
\author{
Yifei Wang\textsuperscript{1}\equalcontrib,
Xiaohan Zhang\textsuperscript{1}\equalcontrib,
Youtao Ding\textsuperscript{1},
Tianlin Li\textsuperscript{2}\corresponding,
Xiaoyu Zhang\textsuperscript{3},
Yida Yang\textsuperscript{4},
Li Pan\textsuperscript{1}
}
\affiliations{
\textsuperscript{1}Shanghai Jiao Tong University, China \\
\textsuperscript{2}Beihang University, China \\
\textsuperscript{3}Nanyang Technological University, Singapore \\
\textsuperscript{4}Tongji University, China
}

\begin{document}

\maketitle

\begin{abstract}
Distributed inference has become an indispensable part of deploying medical models under practical latency, memory, and throughput constraints. Although modern frameworks improve serving efficiency through tensor parallelism, mixed precision, kernel fusion, and multi-device communication, they are generally assumed to preserve the behavior observed during centralized HuggingFace evaluation. This assumption creates an evaluation-deployment mismatch: a model may pass offline evaluation but produce a different output after the execution stack changes. To address this mismatch, we propose a testing framework and an improved, distributed-execution-sensitive medical-model benchmark that evaluates the same checkpoint and input under a centralized HuggingFace reference and matched distributed deployments. Extensive experiments across language, vision, and multimodal medical models show that execution changes can produce measurable output disagreements. Across supported visual settings, the test success rate ranges from 0.21 to 0.43 for single-modality models and from 0.32 to 0.98 for multimodal models. The benchmark is aimed at extending medical-model evaluation from capability and security to evaluation-deployment consistency.
\end{abstract}


\section{Introduction}\label{sec1}

Medical foundation models are widely used for medical image classification, visual question answering and multimodal clinical decision support. Large language models (LLMs)~\cite{touvron2023llama}, vision models and vision-language models (VLMs)~\cite{radford2021learning, li2023blip} underpin modern medical AI. Their evaluation and deployment environments, however, are often different. Offline evaluation commonly uses a centralized single-device HuggingFace configuration, whereas practical serving commonly uses distributed frameworks with tensor parallelism, mixed precision, fused kernels, and multi-device communication to satisfy latency and memory constraints~\cite{aminabadi2022deepspeed, kwon2023efficient, pope2023efficiently}. The deployment workflow therefore relies on a simple but largely untested assumption: a medical image that receives one prediction during evaluation should receive the same prediction after the execution system changes~\cite{aminabadi2022deepspeed, pope2023efficiently, kwon2023efficient}.

\begin{figure}[t]
    \centering
    \includegraphics[width=\linewidth]{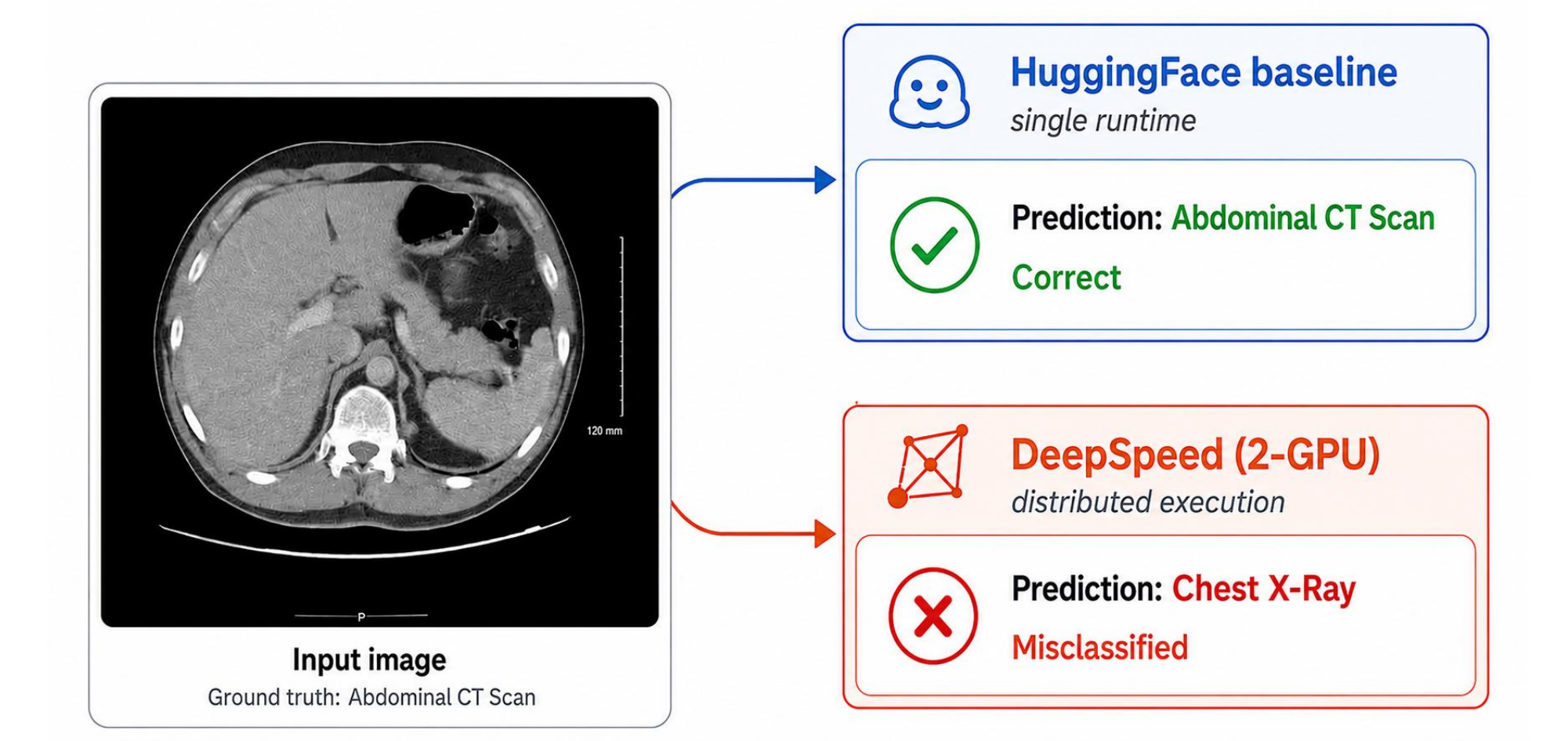}
    \caption{The same abdominal CT image receives different predictions under HuggingFace and two-GPU DeepSpeed inference in our test.}
    \label{fig:intro_medical_discrepancy}
\end{figure}

This assumption exposes an important gap in medical-model evaluation. Existing studies mainly measure factual correctness, reasoning ability, hallucination, and task accuracy on medical question-answering and reasoning benchmarks~\cite{jin2021disease, ben2019question, jin2019pubmedqa, pal2022medmcqa, singhal2025toward, zheng2023judging}; they also examine robustness under harmful or deliberately difficult prompts~\cite{wei2023jailbroken, liu2023jailbreaking}. What they rarely measure is visual deployment consistency: \textbf{whether a model validated on a medical image in one execution environment preserves the same confidence profile and final prediction in another}. Figure~\ref{fig:intro_medical_discrepancy} illustrates the consequence. The same abdominal CT image is classified as Abdominal CT Scan by the HuggingFace reference but as Chest X-Ray by a two-GPU DeepSpeed execution, although the checkpoint and image are unchanged. Thus, a small execution-level difference can become a confidence shift, and a confidence shift can become a different medical prediction. Deployment-induced inconsistency should therefore be tested as an independent reliability dimension rather than hidden inside aggregate accuracy.

To address this evaluation-deployment mismatch, we present a deployment-centered differential testing framework and propose an improved, distributed-execution-sensitive medical-model benchmark. The benchmark follows a paired design: HuggingFace serves as the centralized evaluation reference, and each distributed execution serves as a matched deployment variant. In every comparison, the checkpoint, input semantics, and decision rule are fixed; only the execution stack changes. We instantiate this design with complementary visual and textual test-generation procedures and conduct extensive experiments across language, vision, and multimodal medical models. On the visual track, single-modality models exhibit test success rates from 0.21 to 0.43, whereas multimodal models show larger and more framework-dependent variation, ranging from 0.32 to 0.98 across the evaluated distributed configurations. These results show that deployment disagreement is not limited to language generation, but also affects medical vision and multimodal inference. The benchmark therefore turns an implicit deployment assumption into an explicit evaluation target. Based on its visual track, we further localize the layers that amplify execution differences and apply one targeted mitigation: selectively elevating these critical layers to higher precision while leaving the remaining deployment pipeline unchanged. Our main contributions are as follows:
\begin{itemize}
\item We formulate distributed inference as a deployment-consistency testing problem for medical AI and show that centralized and distributed executions can disagree even when the checkpoint, input, and decision rule are identical.

\item We introduce an execution-aware differential testing framework with symmetric visual and textual test-generation procedures. Based on the visual tests, we further localize divergence-amplifying layers and mitigate disagreements by elevating only these critical layers to higher precision.

\item We propose an improved, distributed-execution-sensitive medical-model benchmark that makes evaluation-deployment consistency an explicit evaluation criterion by pairing centralized HuggingFace evaluation with matched distributed executions.
\end{itemize}

\section{Related Work}

\subsection{Testing and Deployment Reliability}

Machine-learning testing has developed differential testing, metamorphic testing, coverage-guided analysis, fuzzing, and mutation testing to reveal failures in learned systems~\cite{murphy2008properties, dwarakanath2018identifying, pei2017deepxplore, chen2018metamorphic, odena2019tensorfuzz, guo2018dlfuzz, ma2018deepmutation}. Domain-specific studies further examine structural coverage, autonomous-driving perception, and DNN debugging~\cite{xie2019diffchaser, humbatova2020taxonomy, riccio2020testing}. At the system level, CRADLE and Audee show that an identical DL program can behave differently across platforms~\cite{pham2019cradle, guo2020audee, guo2019empirical, wang2024d}; at the language-model level, recent work studies robustness under prompt perturbation and other difficult inputs~\cite{wei2023jailbroken, liu2023jailbreaking, wallace2019universal, zou2023universal}. Our work follows the differential-testing principle but centers the comparison on medical visual behavior: the model, image, and prediction rule remain fixed, while the execution stack changes from a centralized reference to a distributed deployment.

\subsection{Medical Foundation Model Evaluation}

Medical foundation-model evaluation mainly measures capability and semantic safety, including medical image understanding, multimodal perception, question answering, reasoning quality, hallucination, and guideline consistency~\cite{jin2021disease, ben2019question, jin2019pubmedqa, pal2022medmcqa, singhal2023large, singhal2025toward, chen2023huatuogpt, zheng2023judging, ruckert2024rocov2, li2023llava, li2023blip, radford2021learning}. Visual benchmarks evaluate whether a model extracts the correct evidence from an image, while language benchmarks such as MedQA, MedQuAD, PubMedQA, and MedMCQA evaluate whether it produces the correct textual decision. Representative biomedical models include BioBERT, ClinicalBERT, BioGPT, GatorTron, and specialized Qwen variants~\cite{lee2020biobert, alsentzer2019publicly, luo2022biogpt, yang2022large, devlin2019bert, hui2024qwen2}. These efforts answer whether a model is capable under a chosen evaluation environment; they do not fully answer whether the same medical image preserves its confidence and prediction after migration to a different execution environment. We complement them by treating visual deployment consistency as a separate and directly measurable reliability dimension.

\subsection{Numerical Variability in Inference Systems}

Numerical variability is a known property of modern inference systems. Floating-point arithmetic is non-associative, and compiler choices, fused kernels, mixed precision, and parallel reductions can therefore produce different numerical results~\cite{goldberg1991every, higham2002accuracy, kahan1996ieee, demmel2016efficient, ahrens2020algorithms, micikevicius2017mixed}. At the same time, DeepSpeed, PagedAttention, FlashAttention, tensor/model parallelism, quantization, and related techniques improve throughput and memory efficiency precisely by changing execution order, precision, or operator implementation~\cite{aminabadi2022deepspeed, kwon2023efficient, dao2022flashattention, dao2024flashattention, shoeybi2019megatron, rajbhandari2020zero, narayanan2019pipedream, yu2022orca, frantar2022gptq, dettmers2022gpt3}. Recent work further reports inference-engine bugs, distributed-framework issues, hardware-level reliability problems, linear-algebra sensitivity, compiler vulnerabilities, and precision-induced disagreement risks~\cite{liu2025first, yu2025towards, moller2026hardware, moller2025adversarial, chen2025your, wang2026hidden, wang2026trusted}. Our study connects these system-level causes to visual medical outputs: a numerical difference remains harmless when the visual prediction margin is large, but it becomes observable when it changes the leading class or answer associated with the image.

\section{Problem Definition}

\subsection{Definition 1 (Medical Model Execution).}
A medical model $f_\theta$ is a sequence of neural operators
$\langle L_0,L_1,\ldots,L_n\rangle$ parameterized by weights $\theta$.
Given a medical image $x$, or an image-text pair $(x,q)$, the model
transforms the visual evidence layer by layer and produces a class-score
vector or an answer-token distribution. We denote its execution under a
deployment stack $\phi$ by $f_\theta^\phi(x)$, where $\phi$ specifies the
operator kernels, precision policy, tensor partitioning, communication
collectives, attention implementation, and reduction order. A
deterministic decision rule $g(\cdot)$ then converts the numerical output
into an observable visual prediction:
\[
y^\phi(x)=g(f_\theta^\phi(x)).
\]
For image classification, $g(\cdot)$ is an argmax over class probabilities;
for visual question answering, it is answer-option ranking or
autoregressive token selection conditioned on the image. Medical
language-only inference is treated as a complementary instance of the
same execution model.

\subsection{Definition 2 (Deployment-Induced Inconsistency).}
Consider a medical model $f_\theta$, a visual input $x$, a reference
execution stack $\phi_b$, and a distributed or optimized execution stack
$\phi_d$. A deployment-induced inconsistency occurs when
\[
y^{\phi_b}(x) \neq y^{\phi_d}(x),
\]
while the checkpoint, image content, prompt template, and decision policy
remain identical. Thus, the visual evidence is unchanged and the model is
unchanged; only the execution stack differs. This definition is different
from ordinary model error: it asks whether the same medical image receives
a consistent prediction across deployments, not whether either prediction
is clinically correct in isolation.

\subsection{Definition 3 (Deployment-Sensitive Test Case).}
A deployment-sensitive test case is a tuple
$(f_\theta,x,\phi_b,\phi_d)$ for which the same medical image or
image-text pair produces different predictions under the two execution
stacks. For our primary visual and multimodal setting, we generate a
bounded perturbation $\delta$ that preserves the image semantics and the
reference prediction but changes the distributed prediction:
\[
y^{\phi_b}(x+\delta)=y^{\phi_b}(x), \quad
y^{\phi_b}(x+\delta)\neq y^{\phi_d}(x+\delta).
\]
The first condition preserves the behavior validated under the reference
execution; the second condition exposes a deployment-specific prediction
change. For the complementary language setting, we analogously generate
an invisible suffix $s$ such that the visible prompt remains unchanged but
the two executions select different answer options:
\[
y^{\phi_b}(x\oplus s)\neq y^{\phi_d}(x\oplus s).
\]

\subsection{Problem Condition.}
A numerical difference becomes an observable visual inconsistency only
when it crosses a prediction boundary. Distributed execution can introduce
small logit changes through non-associative floating-point computation,
kernel fusion, mixed-precision casting, tensor-parallel reductions,
communication collectives, or memory-efficient attention. When the margin
between the leading visual classes or candidate answers is large, these
changes remain internal; when the margin is small, the same changes can
alter the argmax class or the image-conditioned answer. The key condition
is therefore the interaction between execution-level numerical variation
and a discrete visual decision rule. Following this observation, our
primary test setting uses bounded image perturbations to expose
single-card versus multi-card disagreement, while the language setting
serves as a complementary test of the same condition.

\section{Approach}\label{sec_methods}

We develop two complementary methods, VDST and TDST, to test deployment-induced disagreements in visual and textual medical models, respectively.

\subsection{Visual Deployment-Sensitive Testing}

We first introduce \emph{Visual Deployment-Sensitive Testing} (VDST). VDST formulates visual deployment discrepancy as a paired testing problem in which the same model is executed in two environments: a single-card HuggingFace-style reference and a distributed multi-card deployment. The visual benchmark covers conventional vision models and medical VLMs under multiple framework and GPU configurations; the complete experimental matrix is specified in Section~\ref{subsec:experimental_setting}. Across all visual tasks, the checkpoint, prompt, and decoding rule are fixed, so VDST isolates the execution configuration. Algorithm~\ref{alg:visual_asr} summarizes the VDST generation loop.

\begin{algorithm}[t]
\caption{VDST Test Generation}
\label{alg:visual_asr}
\centering
\begin{algorithmic}[1]
\Require Input image $x$, ground-truth label/answer $y$, single-card model $f_s$, multi-card model $f_m$, perturbation budget $\epsilon$, step size $\alpha$, divergence weight $\lambda$, regularization weight $\beta$, maximum steps $T$
\Ensure Test image $x'$ and success flag
\State Initialize $\delta \sim \mathcal{U}(-\epsilon/4, \epsilon/4)$
\For{$t = 1$ to $T$}
    \State $x' \gets \mathrm{clip}(x + \delta)$
    \State Compute single-card prediction $\hat{y}_s \gets f_s(x')$
    \State Compute multi-card prediction $\hat{y}_m \gets f_m(x')$
    \If{$\hat{y}_s = y$ and $\hat{y}_m \neq y$}
        \State \Return $x'$, success
    \EndIf
    \State Compute $\mathcal{L}_{single}(x', y)$ on the single-card path
    \State Compute $\mathcal{L}_{multi}(x', y)$ on the multi-card path
    \State $\mathcal{L} \gets \mathcal{L}_{single} - \lambda \mathcal{L}_{multi} + \beta \|\delta\|_2^2$
    \State Update $\delta \gets \delta - \alpha \nabla_{\delta}\mathcal{L}$
    \State Project $\delta$ back to the $L_{\infty}$ ball: $\delta \gets \mathrm{clip}(\delta, -\epsilon, \epsilon)$
\EndFor
\State \Return $\mathrm{clip}(x+\delta)$, failure
\end{algorithmic}
\end{algorithm}

For each input image, VDST searches for an additive perturbation $\delta$ within an $L_{\infty}$ budget,
\begin{equation}
x' = \mathrm{clip}(x + \delta), \qquad \|\delta\|_{\infty} \le \epsilon,
\end{equation}
where the implementation uses $\epsilon=0.03$, Adam updates with step size $\alpha=0.01$, and a configured maximum of $T$ optimization steps. The success condition is deliberately asymmetric: the single-card execution must remain correct, whereas the multi-card execution must produce a different prediction. Denoting the single-card predictor by $f_s$, the distributed predictor by $f_m$, and the gold label by $y$, a successful test satisfies
\begin{equation}
f_s(x') = y \quad \text{and} \quad f_m(x') \neq y.
\end{equation}
The reported success rate (SR) is the proportion of evaluated inputs for which this condition is reached. Thus, the objective is not to measure generic model error, but to reveal an error that appears only after the execution configuration changes.

VDST uses the same paired objective for image classification and VQA-style models. For classification, VDST minimizes the single-card cross-entropy and maximizes the multi-card cross-entropy:
\begin{equation}
\mathcal{L}_{vis} = \mathcal{L}_{single}(x', y) - \lambda \mathcal{L}_{multi}(x', y) + \beta \|\delta\|_2^2,
\end{equation}
where $\lambda=1.0$ controls execution divergence and $\beta=0.002$ regularizes the perturbation magnitude. For VLMs, $\mathcal{L}_{single}$ and $\mathcal{L}_{multi}$ are next-token negative log-likelihoods computed against the ground-truth short answer. In both cases, VDST follows three steps: preserve the reference prediction, increase the deployment-side loss, and check whether the two executions disagree.

\subsection{Textual Deployment-Sensitive Testing}

We next introduce \emph{Textual Deployment-Sensitive Testing} (TDST). TDST targets medical multiple-choice inference by keeping the visible question unchanged, appending an invisible suffix, and searching for a suffix that separates the distributed prediction from the HuggingFace reference. The candidate pool contains invisible Unicode characters, mainly variation selectors and formatting code points. TDST retains only characters with stable single-token encodings, so the visible input remains unchanged while the search space remains discrete and reproducible.

We define invisibility operationally rather than typographically. A decoded token is retained only when every character belongs to a Unicode control, format, or separator category, or falls within a standard variation-selector range. Tokens containing null bytes or visible symbols are removed. We also require tokenizer stability: each retained character must map to exactly one token ID and must preserve that mapping under repetition. The admissible test-character pool is

\begin{equation}
\begin{aligned}
\mathcal{V}_{\mathrm{test}}=\{\,c \mid\;&
\forall u\in c,\ 
\mathrm{cat}(u)\in\{\mathrm{Cc},\mathrm{Cf},\mathrm{Zs},\mathrm{Zl},\mathrm{Zp}\}
\lor u\in\mathcal{V}_{\mathrm{sel}},\\
&\mathrm{encode}(c)=[t],\ 
\mathrm{encode}(c\Vert c)=[t,t],\ 
\mathrm{NUL}\notin c
\,\}.
\end{aligned}
\label{eq:4}
\end{equation}

where $\mathrm{cat}(u)$ denotes the Unicode category of code point $u$, $\mathcal{V}_{sel}$ denotes the set of Unicode variation selectors, and $t$ is a stable single-token ID. This definition satisfies two requirements at once: the suffix remains visually imperceptible to human readers, and it remains explicitly represented in the token sequence processed by the model. TDST can therefore test boundary-sensitive logits without changing the apparent medical question.

\begin{figure}[!t]
    \centering
    \includegraphics[width=\columnwidth]{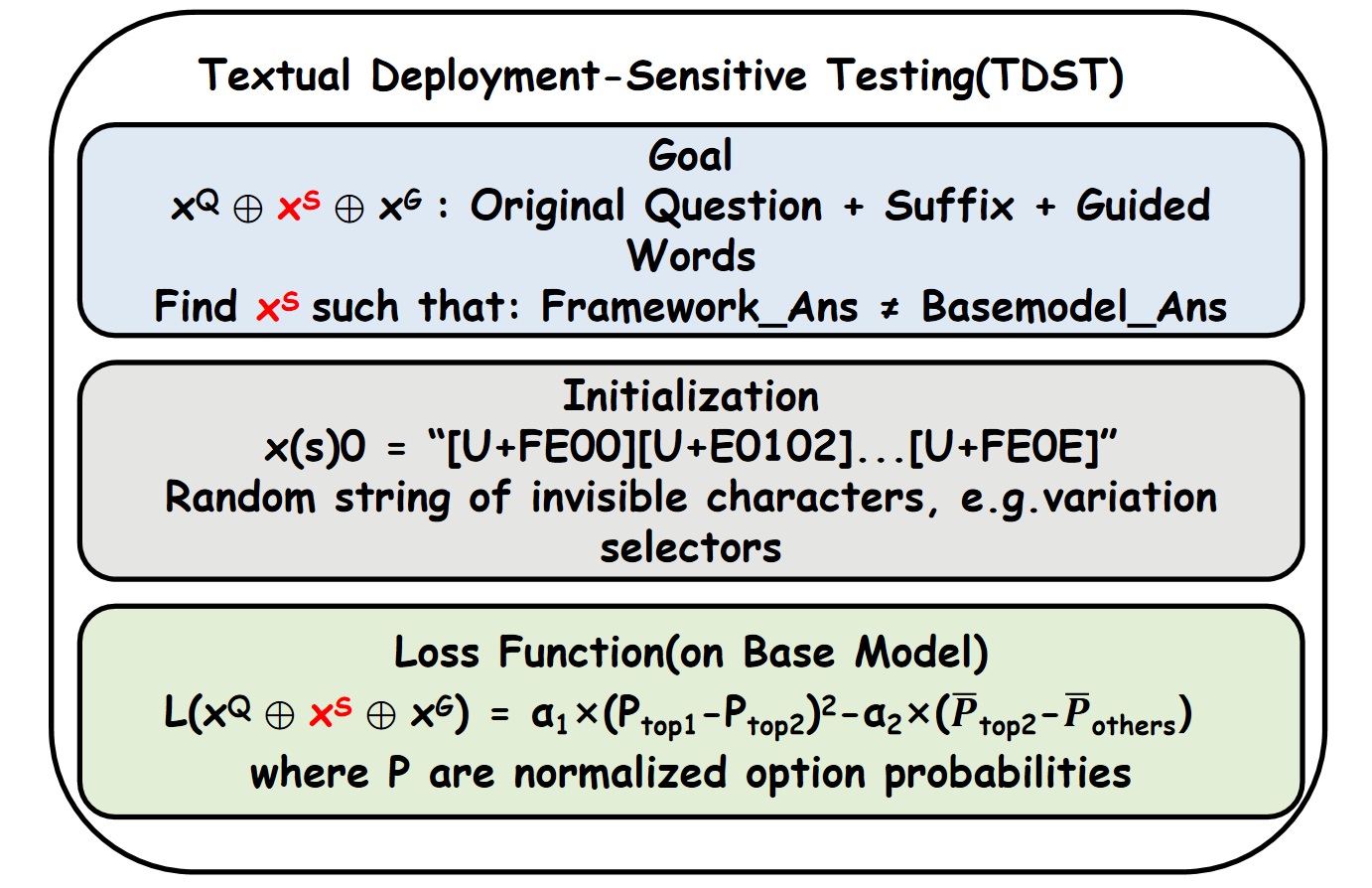}
    \caption{Details of our proposed method TDST.}
    \label{fig:method}
\end{figure}

Let the answer space consist of fixed choice tokens, such as A/B/C/D. TDST first identifies the two highest-probability answers under the reference model. It then reduces the margin between these two answers while preserving their joint dominance over the remaining choices. If $p_{(1)}$ and $p_{(2)}$ are the normalized probabilities of the top-two answers and $\bar{p}_{rest}$ is the mean probability of the other choices, the objective is
\begin{equation}
\mathcal{L}_{txt} = \lambda_{eq}\big(p_{(1)} - p_{(2)}\big)^2 - \lambda_{dom}\bigg(\frac{p_{(1)} + p_{(2)}}{2} - \bar{p}_{rest}\bigg),
\end{equation}
where the first term equalizes the top two and the second keeps them above the distractors. The resulting TDST input moves the prediction toward a narrow boundary, where a small execution-level difference is more likely to change the selected answer.

TDST is gradient-guided but operates in discrete token space. At each iteration, it computes gradients for the current suffix tokens, restricts replacements to the invisible-token pool, selects high-impact candidates from top-$k$ gradient directions, and reruns both reference and distributed executions. TDST stops when the two executions produce different answer letters. Figure~\ref{fig:method} details our TDST, covering its optimization objective, invisible-token initialization, and tailored loss function. The visible medical question remains unchanged, while the tokenizer-preserved suffix steers predictions to the boundary where the two executions yield different answers. VDST works in image space, whereas TDST operates in token space; the two procedures are symmetric in purpose and both reveal deployment-induced disagreement.

\section{Experiments}
\label{sec:results}

We conduct extensive experiments and organize the evaluation around four research questions:

\begin{itemize}
    \item \textbf{RQ1: Natural Discrepancy.} Does the same medical model naturally produce different outputs across deployment frameworks?
    \item \textbf{RQ2: Language Model Testing.} How effectively can TDST expose deployment disagreement in medical language models?
    \item \textbf{RQ3: Vision and Multimodal Testing.} How effectively can VDST expose deployment disagreement in medical vision and multimodal models?
    \item \textbf{RQ4: Localization and Precision Elevation.} Where are numerical differences amplified, and how effectively can critical-layer precision elevation reduce the disagreement?
\end{itemize}

\subsection{Experimental Setting}
\label{subsec:experimental_setting}

\paragraph{Benchmark protocol.}
Our benchmark is designed to measure the mismatch between offline evaluation and distributed deployment. For each test case, HuggingFace provides the centralized evaluation reference, while a distributed framework provides the deployment execution. The checkpoint, input, prompt template, and decision rule are held fixed. Consequently, the comparison changes the execution stack rather than the medical content, allowing the measured disagreement to be attributed to the evaluation-deployment transition.

\paragraph{Language track.}
The language track uses TDST to evaluate four models: Llama-3.2-3B-Instruct, Qwen2.5-3B-Instruct, HuatuoGPT2-7B, and Qwen3-8B on MedMCQA and PubMedQA. We compare HuggingFace with DeepSpeed and vLLM, and all distributed runs use two GPUs. This yields a symmetric design across four models, two datasets, and two deployment frameworks under one fixed GPU configuration.

\paragraph{Vision and multimodal track.}
The visual track uses VDST to evaluate five models under two-GPU and four-GPU execution. ResNet-18 and RCLIP are evaluated on ROCO, while BLIP2-OPT-2.7B, Qwen2.5-VL-7B-Instruct, and Qwen3-VL-8B-Instruct are evaluated on VQA-RAD. The deployment variants include DeepSpeed, FSDP, vLLM, and TensorRT-LLM. ResNet-18 and RCLIP are reported only for DeepSpeed and FSDP because they are not supported by vLLM and TensorRT-LLM in our setup. Table~\ref{tab:cross_framework_all_gpu} summarizes the complete supported model-framework-GPU matrix.

\paragraph{Metrics and mitigation protocol.}
We report Success Rate (SR), the proportion of test inputs for which the benchmark exposes a prediction disagreement, and Avg. Step, the average search steps among successful cases. Localization and mitigation are conducted on the visual track of the benchmark.

\subsection{RQ1: Natural Discrepancy Across Deployment Frameworks}
\label{subsec:rq1}

\begin{figure}[!t]
    \centering
    \includegraphics[width=\columnwidth]{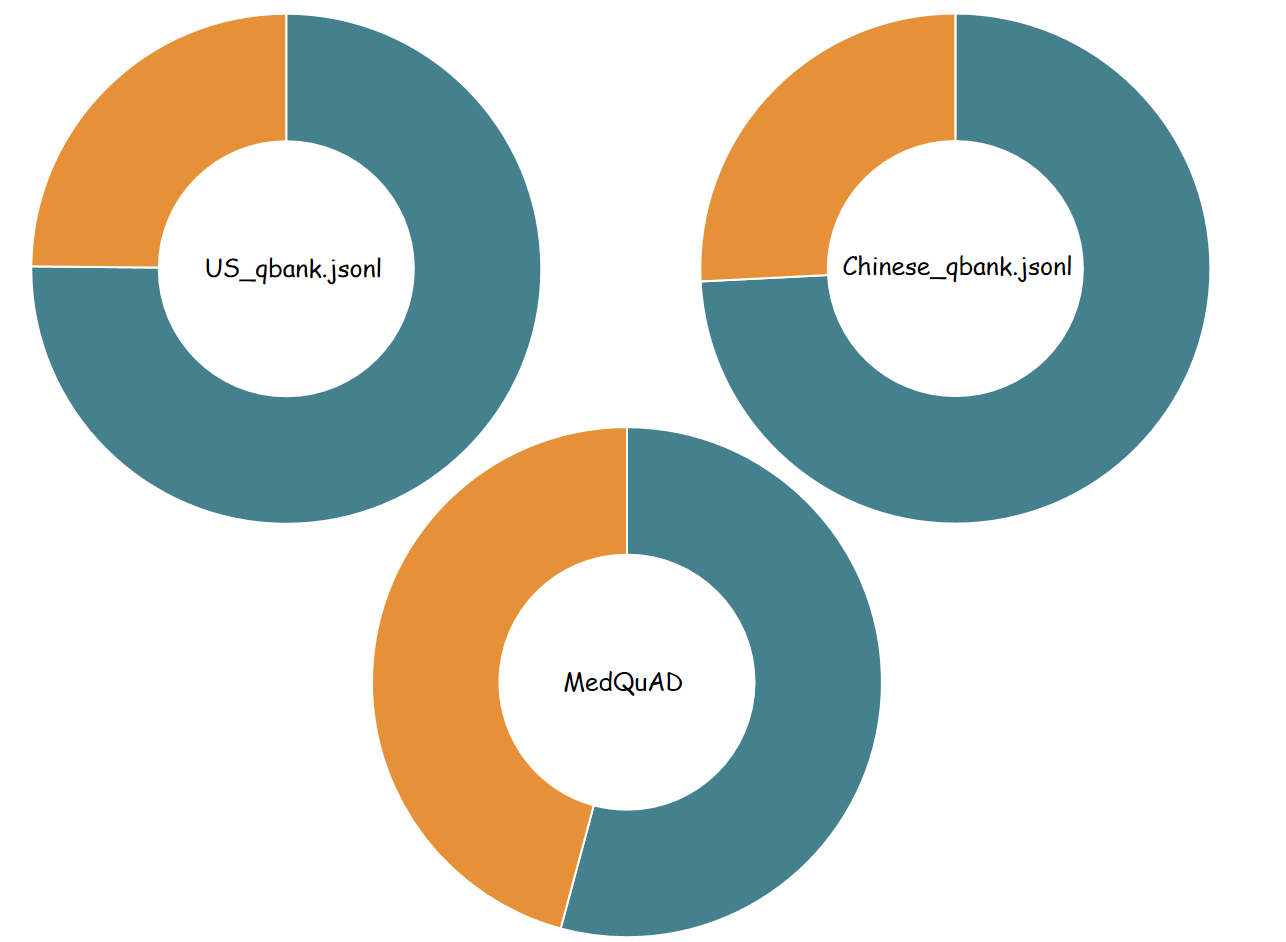}
    \caption{Natural output discrepancy between the standard non-distributed configuration and the two-GPU DeepSpeed configuration on MedQA and MedQuAD.}
    \label{fig:medical_distribution}
\end{figure}

We first measure disagreement on unmodified medical questions, without constructed perturbations or optimization-based test generation. 
We evaluate HuatuoGPT2-7B on three medical QA corpora and compare a standard non-distributed execution with a two-GPU DeepSpeed execution. 
The checkpoint, question, prompt construction, and decoding settings remain unchanged; only the deployment framework changes. 
This paired setup makes the interpretation direct: an output difference is an execution disagreement, not a consequence of different model weights or different question semantics. As summarized in Figure~\ref{fig:medical_distribution}, the natural disagreements are frequent and consistent across datasets. 
On the MedQA-derived US subset, 3,571 of 14,369 samples differ, giving a discrepancy rate of 24.85\%. 
On the MedQA-derived Chinese subset, 8,838 of 34,253 samples differ, giving a similar rate of 25.80\%. 
Together, the two MedQA subsets contain 12,409 disagreements among 48,622 samples, or 25.52\% overall. 
The webMedQA split corresponding to MedQuAD shows a larger difference: 7,517 of 16,407 samples disagree, yielding 45.82\%. 
Thus, the two MedQA subsets exhibit nearly symmetric discrepancy rates, whereas the MedQuAD setting is approximately 20 percentage points higher. In summary, deployment disagreement is not limited to specially generated boundary cases. 
It already appears at scale on ordinary medical questions, and it appears under a simple change from centralized to distributed execution. 
A disagreement does not by itself identify which output is clinically correct, but it does show that the validated behavior is not invariant to deployment. 
RQ1 therefore motivates systematic execution-aware testing before a medical model is released.

\begin{table*}[t]
\centering

\begin{tabular}{
    @{}l
    *{8}{>{\centering\arraybackslash}m{0.85cm}}
    @{}
}
\toprule
\multirow{2}{*}{\textbf{Model}}
& \multicolumn{2}{c}{\textbf{DeepSpeed}}
& \multicolumn{2}{c}{\textbf{FSDP}}
& \multicolumn{2}{c}{\textbf{vLLM}}
& \multicolumn{2}{c}{\textbf{TensorRT-LLM}} \\
\cmidrule(lr){2-3}
\cmidrule(lr){4-5}
\cmidrule(lr){6-7}
\cmidrule(lr){8-9}
& \textbf{SR} & \textbf{Step}
& \textbf{SR} & \textbf{Step}
& \textbf{SR} & \textbf{Step}
& \textbf{SR} & \textbf{Step} \\
\midrule

\multicolumn{9}{c}{\textbf{2-GPU Execution}} \\
\midrule

ResNet-18
& 0.27 & 35.67
& 0.22 & 41.74
& - & -
& - & - \\

\addlinespace[0.2em]
RCLIP
& 0.41 & 100.63
& 0.38 & 86.06
& - & -
& - & - \\

\addlinespace[0.2em]
BLIP2-OPT-2.7B
& 0.51 & 70.90
& 0.47 & 65.81
& 0.98 & 16.72
& 0.96 & 23.65 \\

\addlinespace[0.2em]
Qwen2.5-VL-7B-Instruct
& 0.33 & 72.48
& 0.42 & 31.40
& 0.93 & 22.57
& 0.47 & 156.16 \\

\addlinespace[0.2em]
Qwen3-VL-8B-Instruct
& 0.33 & 66.73
& 0.65 & 46.63
& 0.81 & 40.86
& 0.81 & 29.95 \\

\midrule
\multicolumn{9}{c}{\textbf{4-GPU Execution}} \\
\midrule

ResNet-18
& 0.24 & 37.88
& 0.21 & 44.77
& - & -
& - & - \\

\addlinespace[0.2em]
RCLIP
& 0.43 & 96.81
& 0.42 & 93.79
& - & -
& - & - \\

\addlinespace[0.2em]
BLIP2-OPT-2.7B
& 0.51 & 70.90
& 0.56 & 79.52
& 0.97 & 18.52
& 0.94 & 26.19 \\

\addlinespace[0.2em]
Qwen2.5-VL-7B-Instruct
& 0.32 & 66.09
& 0.45 & 35.73
& 0.92 & 17.70
& 0.48 & 170.14 \\

\addlinespace[0.2em]
Qwen3-VL-8B-Instruct
& 0.36 & 73.64
& 0.66 & 47.56
& 0.85 & 38.47
& 0.82 & 32.57 \\

\bottomrule
\end{tabular}
\caption{Cross-framework evaluation under 2-GPU and 4-GPU execution.
Each framework reports the Success Rate (SR) and Avg. Step.
Note that ResNet and RCLIP are not supported by vLLM and TensorRT-LLM.}
\label{tab:cross_framework_all_gpu}
\end{table*}

\subsection{RQ2: Testing Medical Language Models}
\label{subsec:rq2}

\begin{figure}[!t]
    \centering
    \includegraphics[width=\columnwidth]{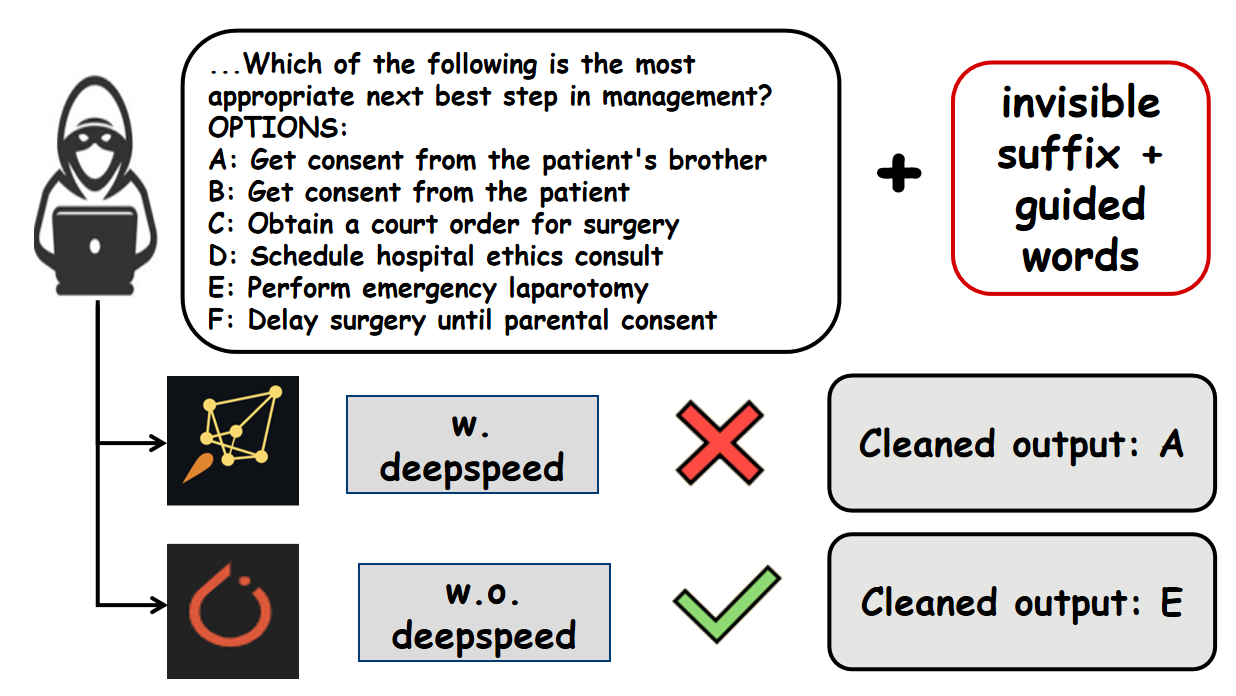}
    \caption{A TDST-generated medical multiple-choice test that produces different answers under HuggingFace and distributed execution.}
    \label{fig:adversarial_medical_mcq_discrepancy}
\end{figure}

\begin{figure}[t]
    \centering
    \includegraphics[width=\columnwidth]{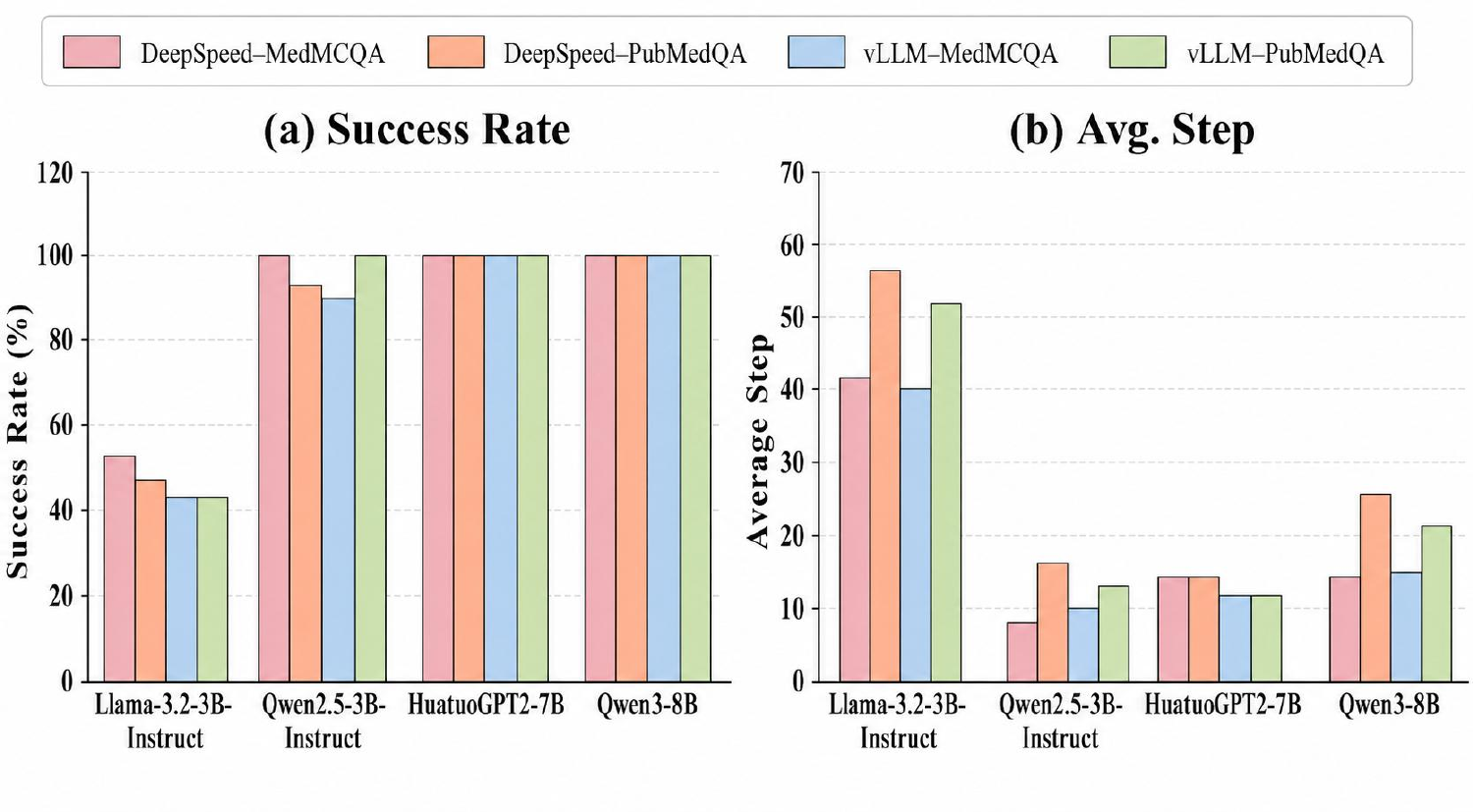}
    \caption{Two-GPU divergence results between HuggingFace and the DeepSpeed and vLLM inference frameworks across four medical language models. Panel (a) reports the success rate of inducing output disagreement, while Panel (b) shows the average number of optimization steps required. Results are evaluated on MedMCQA and PubMedQA.}
    \label{fig:framework_divergence}
\end{figure}

Following the language-track setting in Section~\ref{subsec:experimental_setting}, we next evaluate whether TDST can expose the same inconsistency in medical language models. Figure~\ref{fig:framework_divergence} compares the four models on MedMCQA and PubMedQA under the two-GPU DeepSpeed and vLLM deployments using TDST. The success rate is the proportion of questions for which TDST finds a disagreement within the configured search budget, and Avg. Step is the mean number of steps required for successful cases. The DeepSpeed results show a clear separation between models. Llama-3.2-3B-Instruct reaches 53.33\% SR on MedMCQA and 46.67\% on PubMedQA, requiring 41.13 and 57.14 steps on average, respectively. In contrast, Qwen2.5-3B-Instruct, HuatuoGPT2-7B, and Qwen3-8B all reach 100\% SR on both datasets. Among them, Qwen2.5-3B-Instruct is the easiest to test, requiring only 8.57 steps on MedMCQA and 10.70 steps on PubMedQA; HuatuoGPT2-7B and Qwen3-8B also remain below 15 average steps in all four settings. The vLLM results preserve the same model-level pattern but change the search difficulty. Llama-3.2-3B-Instruct reaches 43.33\% SR on both datasets, with 40.08 and 51.23 average steps. Qwen2.5-3B-Instruct reaches 93.33\% on MedMCQA and 100\% on PubMedQA, while HuatuoGPT2-7B and Qwen3-8B reach 100\% on both datasets. HuatuoGPT2-7B requires only 11.43-12.20 steps, whereas Qwen3-8B requires 21.30-25.60 steps under vLLM despite the same 100\% SR. In summary, TDST succeeds broadly across both frameworks, but SR and search cost are jointly determined by the model, dataset, and runtime rather than by any one factor alone. Figure~\ref{fig:adversarial_medical_mcq_discrepancy} gives a concrete multiple-choice example. 
The invisible suffix narrows the margin between the leading candidate answers, after which HuggingFace and the distributed execution select different options. 
Thus, RQ2 has an affirmative answer: deployment disagreement in medical LLMs can be systematically exposed by TDST, and the required search effort can be measured directly.

\subsection{RQ3: Testing Medical Vision and Multimodal Models}
\label{subsec:rq3}

Following the visual-track setting in Section~\ref{subsec:experimental_setting}, we then evaluate whether VDST exposes deployment disagreement in medical vision and multimodal models. Table~\ref{tab:cross_framework_all_gpu} reports the complete supported comparison under two-GPU and four-GPU execution, with HuggingFace as the centralized reference and DeepSpeed, FSDP, vLLM, and TensorRT-LLM as deployment variants. The results show both breadth and variation. 
For single-modality vision, we use ROCO with ResNet-18 and RCLIP; for multimodal inference, we use VQA-RAD with BLIP2-OPT-2.7B, Qwen2.5-VL-7B-Instruct, and Qwen3-VL-8B-Instruct. 
Across these settings, VDST exposes confidence shifts and prediction disagreements under distributed execution. 
Because the image, checkpoint, prompt, and evaluation protocol are fixed within each comparison, the measured difference is attributable to the serving stack rather than to random data variation. 

Three quantitative patterns are visible in Table~\ref{tab:cross_framework_all_gpu}. First, the single-modality models have comparatively moderate SRs: ResNet-18 ranges from 0.21 to 0.27 and RCLIP from 0.38 to 0.43 across DeepSpeed and FSDP. Second, the multimodal models are substantially more sensitive under several optimized runtimes. Under two-GPU execution, BLIP2-OPT-2.7B reaches 0.98 SR on vLLM and 0.96 on TensorRT-LLM, Qwen2.5-VL-7B-Instruct reaches 0.93 on vLLM, and Qwen3-VL-8B-Instruct reaches 0.81 on both vLLM and TensorRT-LLM. These high SRs are often accompanied by low search cost: BLIP2 requires 16.72 steps on vLLM and 23.65 on TensorRT-LLM. By contrast, Qwen2.5-VL reaches only 0.47 SR on TensorRT-LLM and requires 156.16 steps, showing that successful test generation can be both less frequent and more expensive in a difficult configuration. Third, increasing the execution scale from two GPUs to four GPUs does not remove the pattern. BLIP2 changes only from 0.98 to 0.97 on vLLM and from 0.96 to 0.94 on TensorRT-LLM; Qwen2.5-VL changes from 0.93 to 0.92 on vLLM; and Qwen3-VL changes from 0.81 to 0.85 on vLLM. The exact values shift, but the framework-specific sensitivity remains. RQ3 is therefore answered affirmatively: VDST exposes execution disagreement in both medical vision and multimodal models, while its magnitude and detection cost depend jointly on model architecture, serving framework, and GPU configuration.

\begin{table}[t]
    \centering

    \begingroup
    \begin{tabular*}{\columnwidth}{
        @{\extracolsep{\fill}}
        lcccc
        @{}
    }
        \toprule
        \multirow{2}{*}{\textbf{Model}}
        & \multicolumn{4}{c}{\textbf{Success Rate (SR) $\downarrow$}} \\
        \cmidrule(lr){2-5}
        & \makecell{\textbf{Deep}\\\textbf{Speed}}
        & \textbf{FSDP}
        & \textbf{vLLM}
        & \makecell{\textbf{TensorRT-}\\\textbf{LLM}} \\
        \midrule

        ResNet-18       & 0.00 & 0.00 & -   & -   \\
        RCLIP           & 0.01 & 0.00 & -   & -   \\
        BLIP2-OPT-2.7B  & 0.19 & 0.30 & 0.45 & 0.07 \\
        Qwen2.5-VL-7B   & 0.13 & 0.04 & 0.86 & 0.46 \\
        Qwen3-VL-8B     & 0.24 & 0.38 & 0.17 & 0.01 \\
        \bottomrule
    \end{tabular*}
    \endgroup

        \caption{Cross-framework mitigation performance under 2-GPU execution. We report the post-mitigation success rate (SR) after elevating the localized critical layers to higher precision; lower values indicate stronger mitigation.}
    \label{tab:cross_framework_defense}
\end{table}

\subsection{RQ4: Localization and Precision Elevation}
\label{subsec:rq4}

RQ1-RQ3 show that deployment disagreement exists; RQ4 investigates where numerical differences are amplified and whether a targeted precision change can reduce them. All experiments in this section use the visual track of our benchmark.

\paragraph{Layer-wise localization.}
We compare centralized and distributed executions with the same checkpoint, input, and decoding configuration. These executions may differ numerically because of floating-point non-associativity, mixed precision, kernel fusion, and framework-specific parallel reduction orders. During the forward pass that produces the first generated token, we record the output of every leaf module, since later tokens may already belong to diverged generation trajectories. For layer $i$, let $O_i^{\mathrm{ref}}$ and $O_i^{\mathrm{dist}}$ denote the reference and distributed activations. We measure their Mean Absolute Difference (MAD) as
\begin{equation}
\mu_i =
\frac{1}{N}
\sum_{j=1}^{N}
\left|
O_{i,j}^{\mathrm{ref}}-
O_{i,j}^{\mathrm{dist}}
\right|,
\end{equation}
where $N$ is the number of tensor elements. To identify layers that amplify rather than merely inherit divergence, we define the Relative Divergence Lift (RL):
\begin{equation}
\mathrm{RL}_i =
\frac{
\mu_i-\max_{k<i}\mu_k
}{
\max_{k<i}\mu_k+\varepsilon
}.
\end{equation}
Critical layers are selected using
\begin{equation}
\mathcal{C}
=
\left\{
i \mid
\mathrm{RL}_i >
\mathrm{Percentile}_{95}(\mathrm{RL})
\right\}.
\end{equation}
The detected modules are mainly concentrated around numerically sensitive components such as \texttt{self.rotary\_emb} and \texttt{self.norm}.

\paragraph{Critical-layer precision elevation.}
We then execute only the detected critical modules in higher precision, such as float32, while leaving all other modules and deployment settings unchanged. This targeted strategy limits additional computation and memory compared with globally increasing model precision. Table~\ref{tab:cross_framework_defense} reports the post-mitigation SR under the same two-GPU settings as Table~\ref{tab:cross_framework_all_gpu}. Averaged over the 16 supported model-framework pairs, SR decreases from 0.56 to 0.21. The reduction is especially strong for ResNet-18 and RCLIP, whose SR approaches zero, and for several multimodal settings, such as Qwen3-VL on TensorRT-LLM. However, substantial residual disagreement remains in some configurations, particularly Qwen2.5-VL on vLLM. Overall, RQ4 shows that execution differences are disproportionately amplified by a small set of precision-sensitive modules. Selectively elevating their precision provides an effective minimal-change mitigation, although its effectiveness remains model- and framework-dependent.

\section{Conclusion}
In this paper, we address the mismatch between centralized evaluation and distributed deployment for medical models by introducing a distributed sensitive medical-model benchmark. The benchmark treats evaluation and deployment as a paired system and measures whether model behavior remains consistent when only the execution stack changes. Extensive experiments across language, vision, and multimodal models show that distributed frameworks and GPU configurations can induce observable prediction disagreements, even when the checkpoint, input, and decoding policy are fixed. To further understand and mitigate these discrepancies, we localize the layers that amplify execution differences and selectively improve their numerical precision. This targeted strategy reduces disagreement while avoiding unnecessary changes to the entire model. Overall, our results demonstrate that deployment configurations are not behaviorally neutral and that evaluation-deployment consistency should be treated as an explicit criterion in medical-model assessment.

\small
\bibliography{aaai2027}


\end{document}